\documentclass[]{raa}
\usepackage{graphicx,times}
\usepackage{natbib}
\usepackage{amssymb,amsmath}
\usepackage{multirow}
\usepackage{txfonts}
\bibpunct{(}{)}{;}{a}{}{,}

\usepackage[a4paper=true,dvipdfm=true,pagebackref=true]{hyperref}
\hypersetup{pdftitle = The title of my PDF, pdfauthor = My name, pdfsubject= The subject, pdfkeywords = keyword1 keyword2 keyword3}
\hypersetup{colorlinks = true, linkcolor = green, anchorcolor = red, citecolor = blue, filecolor = red, pagecolor = red, urlcolor = red}

\begin{document}

\title{Spectral Analysis of $\chi$ Class Data of GRS 1915+105 Using TCAF Solution
}

   \setcounter{page}{1}

   \author{Anuvab Banerjee\inst{1}, Ayan Bhattacharjee\inst{1}, Dipak Debnath\inst{2*}, Sandip K. Chakrabarti\inst{2}
}

\institute{\small 1. S. N. Bose National Centre for Basic Sciences, Block-JD, Sector-3, Salt Lake, Kolkata, 700106, India\\
   2. Indian Centre for Space Physics, 43 Chalantika, Garia St. Rd., Kolkata, 700084, India\\ %{\it e-mail: dipakcsp@gmail.com; sandip@csp.res.in}\\
 {\it e-mail: anuvab.banerjee@bose.res.in; ayan12@bose.res.in; dipakcsp@gmail.com; sandip@csp.res.in}\\
 \vs\no
   {\small Received:~~2020~~Feb 20; Accepted:~~2020~~Jun 4}
}

%\begin{abstract}
\abstract{
The class variable source GRS 1915+105 exhibits a wide range of time variabilities in time scales of a few seconds to a few days. Depending
on the count rates in different energy bands and the nature of the conventional color-color diagram, the variabilities were classified into
sixteen classes, which were later sequenced in ascending order of Comptonization Efficiency (CE). The CE is the ratio of the phenomenological 
power-law and disk blackbody model fitted spectral photon fluxes. For a better understanding of the flow dynamics in one of the variability classes, 
namely $\chi$, we fitted spectra with the physical two-component advective flow (TCAF) model. In $\chi$ class, X-ray flux is steady with no 
significant variation, however, various $\chi$ subclasses are observed at different X-ray flux and variations of count rates across different 
$\chi$ subclasses must be linked to the variation of flow parameters such as the accretion rates, be it the Keplerian disc rate and/or the 
low angular momentum halo rate. We find that in the $\chi_{2,4}$ classes, which are reportedly devoid of significant outflows, the spectra 
could be fitted well using TCAF solution alone. In the $\chi_{1,3}$ class, which are always linked with outflows, a cutoff power-law model 
is needed in addition to the TCAF solution. At the same time, the normalization required by this model along with the variation of photon index 
and exponential roll-off factor provides us with the information on the relative dominance of the outflow in the latter two classes. 
TCAF fit also provides us with the size and location of the Compton cloud along with its optical depth. Thus by fitting with TCAF, 
a physical understanding of the flow geometry in different $\chi$ classes of GRS 1915+105 has been obtained.
%\end{abstract}
%\end{keywords}
}
 \authorrunning{Banerjee, Bhattacharjee, Debnath {{\&}} Chakrabarti}
   \titlerunning{Properties of GRS 1915+105 $\chi$ Class} 
   \maketitle

\section{Introduction}
It has been noted right since the discovery of the Galactic black hole X-ray binary (BHXB) GRS 1915+105 that it exhibits persistent brightness 
(Remillard \& McClintock, 2006). A complex timing and spectral variability pattern over a wide span of time-scales from a few seconds to 
a few days as revealed through continuous X-ray monitoring is also an important characteristics of this source (Greiner et al., 1996; 
Morgan et al., 1997; Klein-Wolt et al., 2002). This variability pattern was grouped into several classes on the basis of the ratio of 
photon count rates in different energy bands (hardness ratio) and the color-color diagrams obtained from these ratios (Yadav et al., 1999; 
Muno et al., 1999; Belloni et al., 2000). The $\chi$ classes are devoid of strong temporal and spectral variabilities. The subclasses
$\chi_{1,3}$ are associated with strong radio jets (Naik \& Rao, 2000; Vadawale et al., 2001, 2003) while the subclasses $\chi_{2,4}$ 
do not show such a behaviour.

Since the nature of the hardness ratio or color-color diagram depends on the choice of energy bands for the soft and hard photons, and therefore, 
on the mass of the black hole (BH), an alternate mass independent description of these variabilities classes was given (Pal, Chakrabarti \& Nandi, 2013) 
in terms of the Comptonization efficiency (CE) which, in reality, is a dynamic hardness ratio, given by the instantaneous ratio of the number of 
photons under the power-law ($N_p$) and the number of photons under multi-color black body ($N_b$) component of the composite spectra. 
When CE (=$N_p/N_b$) is arranged in ascending order, the classes appear to be exhibited in the same order. Since $N_p$ is nothing but a 
fraction of $N_b$ which are intercepted by the Compton cloud, CE also gives a idea of the flow configuration - a large CE would mean 
a large Compton cloud, which will necessarily yield a hard state. In this sequence, $\chi$ naturally appeared at the end, where the 
spectra are the hardest. In this scheme, similar `looking' classes as far as the light curves go, have similar CEs and the sequence 
is also the same for other objects (e.g., IGR 17091-3624) which exhibit variability classes (Pal \& Chakrabarti, 2015). 

Successful spectral fits and extraction of accretion flow parameters were achieved for a large number of transient black hole sources 
(Debnath et al. 2014, 2015a,b, 2017, 2018, 2020; Mondal et al., 2014, 2016; Chatterjee et al., 2016, 2019; Jana et al., 2016, 2017, 2020a,b; 
Molla et al., 2016; Bhattacharjee et al., 2017; Shang et al. 2019; Chatterjee et al. 2020) with a two component flow solution which is a 
natural outcome of a viscous transonic flow (Chakrabarti, 1995, 1997 and references therein). Recently, the same paradigm has also been 
applied in case of persistent sources and weakly magnetized neutron stars with requisite modifications (Banerjee et al., 2019; Bhattacharjee 
and Chakrabarti, 2017, 2019; Bhattacharjee, 2018). This so-called Two Component Advective Flow (TCAF) solution envisages that due to vertical 
gradient of viscous processes, the injected low angular momentum matter would disaggregate into an equatorial plane based Keplerian disc emitting 
soft X-rays surrounded by a low angular momentum halo forming a centrifugal barrier close to the black hole and emitting Comptonized hard X-rays. 
The centrifugal barrier could be so strong that the flow piles up against it forming a standing shock and the subsonic post-shock region 
is the CENtrifugal pressure supported BOundary Layer or CENBOL. This acts like a boundary layer and is also responsible for supplying matter 
to form outflows (Chakrabarti, 1999). Beyond the shock location, the Keplerian and sub-Keplerian components are mixed and therefore, shock location acts as the truncation radius in TCAF paradigm. Giri \& Chakrabarti (2013) established through numerical simulations that the TCAF solution is a 
stable configuration. In the harder states, the CENBOL cannot be cooled by inverse Comptonization  (e.g., Sunyaev \& Titarchuk, 1980, 1985) 
due to low disc accretion rate and thus outflow is produced. In soft states, the CENBOL is completely cooled down and collapses due to 
high Keplerian disc rate and no outflow can form. This is also established by detailed hydrodynamic simulations with Compton processes 
(Garain et al. 2012). It has been shown also that the ratio of the outflow rate to inflow rate is highest if the shock compression ratio 
is moderate (Chakrabarti, 1999) and this outflow takes part in Comptonization and plays a major role in deciding the nature of the 
light curves of GRS 1915+105 (Chakrabarti \& Manickam, 2000).

In the literature, X-ray spectral properties of GRS 1915+105 has been studied by several authors over the years {(Fender \& Belloni, 2004; 
Remillard \& McClintock, 2006; McClintock \& Remillard, 2006; Pal et al., 2013, 2015; Peris et al., 2016). It has also been a key source in 
understanding the disk-jet connection in the context of X-ray binaries (Vadawale et al., 2003; Fender, Belloni \& Gallo, 2004). The radio 
flares have been subdivided into two categories on the basis of the properties of accretion disk evolution: (1) Persistent radio flare where 
the accretion disk remain steady. In this case inflow and outflow are in equilibrium and there is no sudden change in either soft or 
hard component. (2) Other flares which are associated with changes in the accretion disk. Here the hard and soft components are affected 
and the object proceeds towards a state transition (Yadav 2001, 2006). Pal et al., (2013, 2015) studied mainly with the spectral fits using 
phenomenological power-law (PL) and disk blackbody (DBB) models and also to obtain the hardness of the object in different timing classes. 
However, to the best of our knowledge, no work has been executed which compute the accretion rates and properties of the Compton cloud for 
any of the variables classes. Just like TCAF model is employed to study the accretion flow dynamics of the outbursting sources 
(Debnath et al. 2014, 2015a), the same fitting procedure could be used for GRS 1915+105 as well to study the dependence of class behaviours 
on flow parameters. 

In the present paper, we concentrate on the analysis of the $\chi$ class data of RXTE satellite to determine the flow parameters. 
In the next Section, we first give a theoretical background on the flow configuration as expected from the TCAF solution and how winds 
might be formed which affect the spectra in this class of solutions. In \S 3, we analyze the data. In \S 4, we present our results. 
Finally, in \S 5, we give the concluding remarks.

\section{Theoretical background}

It is normally assumed that a Keplerian flow is supplied at the outer edge of a black hole accretion disc by the companion 
(Shakura \& Sunyaev, 1973). However, observations of several binary systems clearly suggest that two components are necessary to 
explain the timing properties as proposed by Chakrabarti in 1997 (Chakrabarti 1997; Smith et al., 2002; Wu et al., 2002; 
Ghosh \& Chakrabarti 2019). Subsequently, it was shown that the jets and outflows are launched at the CENBOL and the outflow rate 
depends on the compression ratio of the shock which is responsible to form the CENBOL (Chakrabarti 1999, hereafter C99). The nature 
of the two component inflow configurations across the spectral states and when the outflows are expected to be highest have been 
discussed in Chakrabarti et al. (2000), Chakrabarti \& Nandi (2000).

In Fig. 1(a,b), adopted from Chakrabarti et al. (2000), we show the two possible flow configurations leading to $\chi_{2,4}$ and $\chi_{1,3}$ 
temporal states respectively. In Fig. 1(a), we have shown the standard flow configuration around the black hole consisting of Keplerian and 
sub-Keplerian matter both.  If the shock conditions are satisfied and the viscosity is so small that the Keplerian disc rate is still low, 
the outflow may form close to the black hole. Until the optical depth of the sonic sphere at base of the outflow is low and thus the outflow 
is steady. This configuration is depicted in Fig. 1(b) (also see, C99). Initially, as the shock strength rises from unity, the outflow rate 
would also rise (C99). When the optical depth of the sonic sphere crosses unity, the outflow will separate out as a blobby jet component and 
the cooler matter returns back to the inflow (Chakrabarti \& Nandi, 2000). This creates the so-called `burst-on' state. When the return flow 
is drained out, a `burst-off' state is created. A further rise in the shock strength would reduce the outflows in comparison to inflows. 
When the viscosity is high enough to enhance the Keplerian disc rate, it cools the CENBOL, quenches the outflow, and a soft state is created.

\begin{figure}[htb]
    \centering
\begin{subfigure}{0.40\textwidth}
  \includegraphics[width=\linewidth]{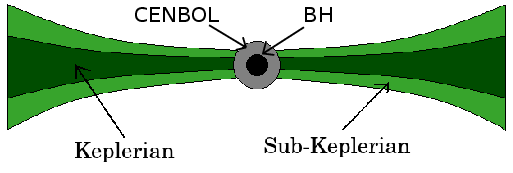}
  \caption{}%(a)}
  \label{fig:1a}
\end{subfigure}\hfil
\begin{subfigure}{0.40\textwidth}
  \includegraphics[width=\linewidth]{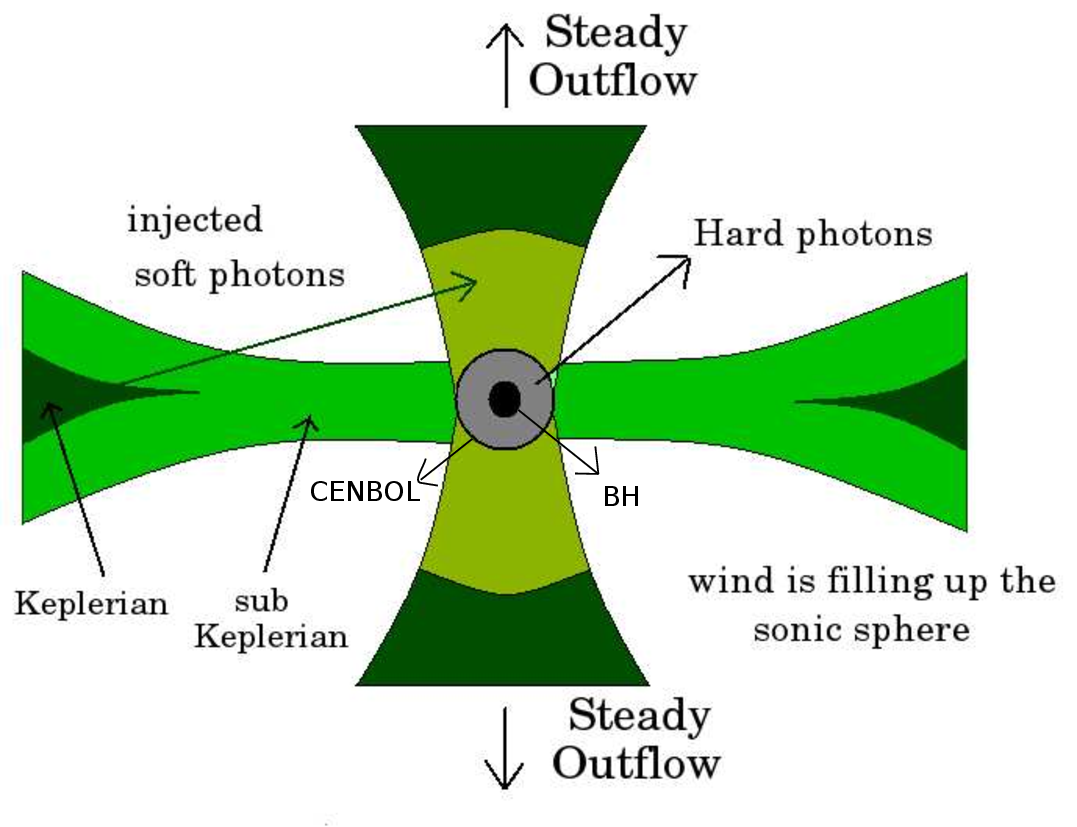}
  \caption{}%(b)}
  \label{fig:1b}
\end{subfigure}\hfil

\caption{(a) The accretion flow consisting of Keplerian and sub-Keplerian flow. The relative abundance of them dictates the hardness 
and softness of the state. This configuration corresponds to $\chi_{2,4}$ class. (b) Shock produced due to satisfaction of Rankine-Hugoniot 
conditions. In both of the figures, the grey shaded region around the black hole represents the CENBOL. Outflow generated and a steady outflow rate is maintained because the average optical depth within the sonic sphere is smaller 
than unity. This corresponds to $\chi_{1,3}$ class (adopted from Chakrabarti et al., 2000).}
\label{fig1}
\end{figure}

In the recent time as TCAF paradigm has been well studied in case of the outbursting sources, we ventured to extend our spectral studies to the 
class variable source GRS 1915+105 as well to obtain a possible picture of accretion flow dynamics around such objects. Muno et al. (1999) and 
McClintock et al. (2006) reported that classes with steady fluxes from GRS~1915+105 often resemble hard states of persistent sources. In the 
spectral and temporal study of this object by Rao et al. (2000) to pin down the source of the hard photons to be the CENBOL, the spectra of 
$\chi_3$ class were also fitted using combined disk blackbody (diskbb), CompST and cut-off power-law (diskbb+CompST+cutoffpower-law) models. 
Since for $\chi$ classes the fluxes are almost steady, we were motivated to do the spectral analysis using TCAF model on $\chi$ classes. 
Since in TCAF, the contribution from disk and Comptonization from the CENBOL are self-consistently accounted for by solving radiative transfer 
equation, there is no requirement for separate models like diskbb and CompST. However, to account for the additional Comptonization from 
the outflow, other models like cutoff~power-law (cutoffpl) might be required. 

According to the reports in literature, mass of the source is believed to be well established. It had been first estimated to be $14 \pm 4 M_\odot$ (Greiner et al., 2001). Investigating the optical counterpart, the mass of the donor was found to be $\sim 1.2 ~ M_\odot$, implying the source to be low-mass X-ray binary. A further estimate of $12.4_{-1.8}^{+2.0} ~ M_\odot$ was found by trigonometric parallax measurement (Reid et al., 2014). However, mass is the intrinsic property of the black hole and not a flow parameter. Therefore, for the purpose of our analysis, we have kept the mass of the black hole frozen at $14M_\odot$, while fitting spectra with the TCAF model fits file. 
 
\section{Data selection and the method of analysis}

For the spectral analysis, RXTE science data from NASA HEASARC data archive are used. We consider one each of $\chi_1, \chi_2, \chi_3, \chi_4$ class 
data as reported in Pal et al. (2013) and continuous observation span is divided into several slices to carry out spectral analysis over each one of 
the segments separately in order to have a better statistics on fitted parameters. HEASARC's spectra generating software package HEASOFT version 
HEADAS 6.18 and XSPEC version 12.8.2 has been used for the extraction and analysis of the spectra. For the generation of `.pha' files and fitting 
of the spectra using TCAF solution, the procedure described by Debnath et al. (2013, 2014) has been employed. The \textit{standard2} mode Science 
Data of PCA instrument was used for spectral analysis. For each observational ID, the spectrum was extracted from all the Xenon layers of PCU2 
containing 128 channels, without any grouping. The PCA background was extracted by `runpcabackest' command and then using the latest bright-source 
background model. In order to take care of the South Atlantic Anomaly (SAA), PCA SAA History file was incorporated and the data acquired during the 
SAA passage and for elevation less than $10^\circ$ and offset less than 0.02 has been excluded. The $2.5-25$~keV background subtracted spectra were 
fitted by TCAF based additive model fits file. We have checked that the spectral fitting up to higher energy does not change the qualitative conclusions 
that we have made, hence we have confined ourselves in the aforementioned energy domain. The `err' command was used to determine the 90$\%$ confidence 
error values of the model fitted parameters. 

In order to take care of the interstellar absorption, the multiplicative model \textit{phabs} was employed. While in Muno et al. (1999) the hydrogen 
column density was kept fixed at $6.0 \times 10^{22}$ atoms$/cm^2$, it was kept frozen at $5.0 \times 10^{22}$ atoms$/cm^2$ in the process of fitting 
the steady state spectra by other observers (Lee et al., 2002; Peris et al., 2016). The column density was also kept in the range $(4.5-7.0) \times 10^{22}$ 
atoms/$cm^2$ by Peris et al. (2016), but no significant departure of the fitted parameters were noticed. While analyzing the low/hard state of GRS 1915+105 to measure the spin, Blum et al. (2009) considered the absorption column density to be $4.0 \times 10^{22}$. On the other hand, while measuring the correlation between disk parameters and superluminal jet parameters, $N_H$ had been considered to be at a higher value of $(10.0-15.0) \times 10^{22}$ (Yadav 2006). Throughout our analysis, we have found $N_H$ to be in the range $(4.5-7.5) \times 10^{22}$.

The PCA data IDs chosen for the analysis of $\chi_2$ and $\chi_4$ class data respectively are 20402-01-16-00 (MJD=50501) and 10408-01-33-00 (MJD=50333). 
For $\chi_1$ and $\chi_3$ classes, we used PCA data of observation IDS 10408-01-23-00 (MJD=50278) and 20402-01-50-00 (MJD=50735) respectively. 
The $\chi_{2,4}$ class data are fitted with the TCAF solution based model \textit{fits} file only, for which four input parameters other than normalization (N)
are supplied (as mass of the black hole is kept frozen at $14\dot{M}_\odot$): (i) Keplerian disk rate ($\dot{m}_d$ in $\dot{M}_{Edd}$), (ii) sub-Keplerian halo 
rate ($\dot{m}_h$ in $\dot{M}_{Edd}$), (iii) location of the shock front ($X_s$ in Schwarzschild radius $r_s=2GM/c^2$), (iv) compression ratio R (=$\rho_{+}/\rho_{-}$, 
i.e. ratio of post-shock to pre-shock density). In TCAF model, two auxiliary parameters, namely, the mass of the black hole (in solar mass $M_\odot$) 
and normalization (N) (it properly scales up or down the entire model spectra to match the observed spectra) are generally found to be constant. $N$ depends on 
the intrinsic source parameters: mass of the black hole, distance and disk inclination angle. So, in general $N$ does not vary for a particular source (in observed 
with the same satellite instrument). But, one may require higher $N$ values to fit spectra if there is significant effects of jet or other physical processes, whose 
effects are not included in the present model fits file. To obtain best model fits, a Gaussian emission line of peak energy at around $6.5$~keV is used 
to take care of the iron line emission.

Although $\chi_{2,4}$ are radio quiet, $\chi_{1,3}$ classes are radio loud i.e., there are significant radiation from jets or outflows. In later two  
classes, the TCAF model was not sufficient to fit spectra due to the presence of outflows which could be emitting non-thermal X-rays (see C99, Jana et al. 2017). 
Thus, to obtained best fit in these two classes, we used an additional cutoffpl model. The cutoffpl profile is given by the equation 
$A(E) = KE^{-\alpha}\exp(-E/\beta)$, where $\alpha$ is the power-law photon index, $\beta$ is the exponential roll-off in keV and $K$ is the normalization 
in the unit of photons/keV/$cm^{2}$/s. The strength of $\alpha$ and $\beta$ parameters will indicate relative dominance of the outflow in $\chi_1$ and $\chi_3$ 
class and could throw some light regarding the accretion flow dynamics around the black hole. In order to separate out the contributions from the TCAF system 
and the outflow system, we allowed the Normalization parameter $N$ to vary. Variation of the peak flux would be dependent on the variation of the accretion rates, 
and therefore, any error in determining the accretion rates would be reflected in the error of normalization. Even observational data quality may cause its 
fluctuation. Keeping these in mind, we first determined the energy range up to which the spectra could be fitted using TCAF solution alone. This allows us to 
determine the average normalization which was kept frozen to obtain spectral fits in the $2.5-25$~keV energy range. Our procedure enabled us to separate the 
outflow contribution in the spectra using the method of Jana et al. (2017).

\section{Results}

\subsection{Analysis of $\chi_2$ and $\chi_4$ classes}

A comparison of the spectral natures of $\chi_{2}$ class (Obs. Id. 20402-01-16-00) and $\chi_4$ class (Obs. Id. 10408-01-33-00) imply that  the energy flux is higher in $\chi_4$ across the entire range of the spectra (Figure 2). Since the spectral features are ultimately governed 
by the accretion flow dynamics and consequently by the flow parameters, the different features of the observed spectra of $\chi_{2,4}$ classes 
should be corroborated by the spectral fitted parameters.

\begin{figure}[h!]
\vskip 0.5cm
\begin{center}
\includegraphics[scale=0.6,angle=0,width=7.5truecm]{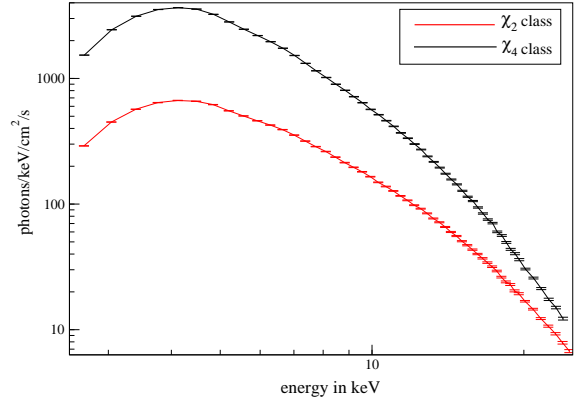}
\caption{Comparison of the observed spectra for $\chi_2$ and $\chi_4$ class data. The low energy flux in case of $\chi_4$ class is greater 
than that of $\chi_2$ class.}
\label{fig2}
\end{center}
\end{figure}

\begin{table}
\small

\vskip0.2cm

{{\bf Table 1:} TCAF Model Fitted Parameters for $\chi_2$ and $\chi_4$ class data in the 2.5-25 keV energy band with the normalization 
as free. Variations of TCAF flow parameters, viz. disk accretion rate ($\dot{m}_d$), halo rate ($\dot{m}_h$), shock location ($X_s$), 
the shock strength (R) and model normalization (N) with the chosen time segments along with the respective error bars are listed. 
Reduced $\chi^2$ of the spectral fits are listed in the last column.}
{\centerline{}}
\begin{center}
\begin{tabular}{c c c c c c c c c}
\hline
\hline
\rm Variability  & $ \rm \dot{m}_d $ & $ \dot{m}_h $ & $ X_s $ & $ R $ &
$ N $ & $\chi^2/dof$ \\
 class & (in $\dot{M}_{Edd}$) & (in $\dot{M}_{Edd}$) & (in $M_\odot$) &  (in $r_s$) & & & & \\

 \hline
 \vspace{0.2cm}
 $\chi_2$ &   $0.797_{-0.02}^{+0.02}$ &  $0.161_{-0.001}^{+0.001}$ & $47.302_{-0.07}^{+0.07}$ & $1.233_{-0.001}^{+0.001}$ & $22.125_{-0.015}^{+0.016}$ & $55.08/45$  \\
 \vspace{0.2cm}
$\chi_4$ & $4.470_{-0.06}^{+0.06}$ & $1.132_{-0.001}^{+0.001}$ & $38.635_{-0.06}^{+0.06}$ & $1.080_{-0.001}^{+0.001}$ & $23.448_{-0.092}^{+0.093}$ & $46.84/44$ \\
  
\hline
\end{tabular}
\end{center}
\end{table}

In case of $\chi_2$ class, the continuous observation spanning over $3000$~seconds has been selected and fitted using TCAF+Gaussian model. 
The flow parameters, namely the disk accretion rate ($\dot{m}_d$), the halo rate ($\dot{m}_h$), shock location $(X_s)$, strength of the 
shock ($R$) have been determined. The same task has been done in case of $\chi_4$ class data as well over the 800 second continuous 
observation span. The spectral fitted parameters for the two classes are mentioned in Table 1. 

\par 

It is evident that the disk rate in case of $\chi_4$ class, the disk accretion rate is significantly larger compared to that of $\chi_2$ class. 
This is in agreement with the higher low energy flux of $\chi_4$ class data. However, since $\chi_{2,4}$ classes are devoid of any outflow, 
the normalization remain almost the same ($\sim 20$). In case of $\chi_4$ class, the enhancement of disk rate makes the shock location to move 
inwards. The data to model ratio for the fitted spectrum in $\chi_2$ class is provided in figure 5(a). In figure 6(a), we provide the 66-90-99$\%$ 
contours for the fitted parameters $\dot{m}_d$ and $\dot{m}_h$ in case of $\chi_2$ class.

\subsection{Analysis on $\chi_1$ and $\chi_3$ class data}

We now summarize our findings from the spectral analysis of $\chi_1$ and $\chi_3$ class data. As mentioned earlier, unlike $\chi_{2,4}$ class, 
substantial activity in the radio domain has been observed which implies the presence of an outflow. In order to account for the additional 
Comptonization from the base of the outflow, we used cutoffpl model in addition to TCAF solution to fit the spectra in $2.5-25$~keV 
energy range. A comparison of the nature of the observed spectra of $\chi_1$ class (Obs Id. 10408-01-23-00) and $\chi_3$ class (Obs Id. 20402-01-50-00) 
implies that the flux at lower energy is more in case of $\chi_1$ class.
\par

Because of the presence of an outflow, the spectral fitting solely with TCAF could not be accomplished over the entire range, but over only 
a restricted energy band. The relative hardness of $\chi_3$ implies the difference in radio dominance in $\chi_1$ and $\chi_3$ classes. 
For this reason, the energy range of feasible TCAF fitting in these two classes were different. 
\par 
Our objective was also to see whether contribution from TCAF and cutoffpl could be segregated. In that case, there would be a possibility 
that the accretion and outflow behaviours would be explainable independently without interference of one on another. For that purpose, after fitting 
the spectra using TCAF in the smaller energy range, the obtained normalization were kept frozen in the TCAF+cutoffpl fitting over the entire 
$2.5-25$~keV energy range.

\begin{figure}
\vskip 0.4cm
\begin{center}
\includegraphics[scale=0.6,angle=0,width=7.5truecm]{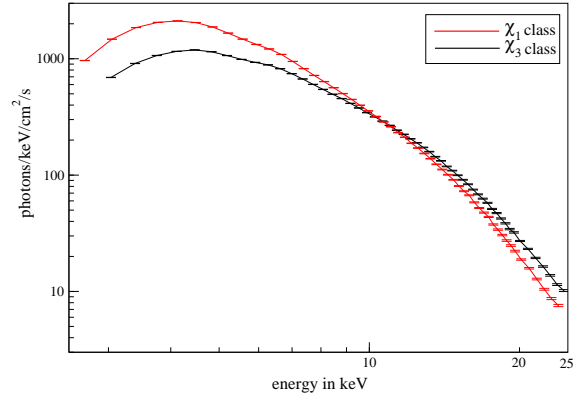}
\caption{Comparison of the spectra for $\chi_1$ and $\chi_3$ class. The energy flux at lower energy is greater in case of $\chi_1$ class. 
The relative hardness of $\chi_3$ class is apparent from the spectral slopes.}
\end{center}
\label{fig3}
\end{figure}

\begin{table}
\vskip 0.0cm
\addtolength{\tabcolsep}{-4.5pt}
\small
\vskip0.2cm

{{\bf Table 2:} TCAF Model Fitted Parameters of $\chi_1$ and $\chi_3$ class PCA spectra in the 2.5-25 keV energy band with model 
normalization kept free. We have listed the variations of TCAF parameters, viz. disk accretion rate ($\dot{m}_d$), halo rate ($\dot{m}_h$), 
shock location ($X_s$), the shock strength (R) and model normalization (N) and the cutoffpl parameters with the chosen time 
segments along with the respective error bars. The reduced $\chi^2$ of the spectral fits are also listed in the last column.}
{\centerline{}}
\begin{center}
\begin{tabular}{c@{\hskip 0.1in} c@{\hskip 0.1in} c@{\hskip 0.1in} c@{\hskip 0.1in} c@{\hskip 0.1in} c@{\hskip 0.1in} c@{\hskip 0.1in} c@{\hskip 0.1in} c@{\hskip 0.1in} c@{\hskip 0.1in} c@{\hskip 0.1in} c}
%\begin{small}
\hline
\hline
\rm  variability & $ \rm \dot{m}_d $ & $ \dot{m}_h $ & $ X_s $ & $ R $ & 
$ N $ & $\alpha$ & $\beta$ & cutoffpl & $\chi^2/dof$ \\
 class & (in $\dot{M}_{Edd}$) & (in $\dot{M}_{Edd}$) &  (in $r_s$) & & & & (in keV) & Norm  & \\
 \hline
\vspace{0.2cm}
            $\chi_1$ &    $1.750_{-0.02}^{+0.02}$ & $0.362_{-0.001}^{+0.001}$ & $27.5_{-0.27}^{+0.31}$ & $1.052_{-0.001}^{+0.001}$ & $17.56_{-0.16}^{+0.16}$ & $2.082_{-0.002}^{+0.003}$ &  $15.9_{-0.09}^{+0.09}$ & $9.10_{-0.038}^{+0.039}$ & $35.92/41$\\
    \vspace{0.2cm}         
           $\chi_3$ &    $1.131_{-0.01}^{+0.01}$ & $0.182_{-0.002}^{+0.003}$ & $20.1_{-0.02}^{+0.03}$ & $1.438_{-0.002}^{+0.002}$ & $2.85_{-0.06}^{+0.06}$ & $1.563_{-0.01}^{+0.01}$ & $15.7_{-0.08}^{+0.08}$ & $3.23_{-0.003}^{+0.003}$ & $42.63/40$\\
           
 \hline
 \vspace{0.2cm}
 \end{tabular}
 
\end{center}

\end{table}

\begin{table}

\small
\vskip0.2cm

{{\bf Table 3:} TCAF Model Fitted Parameters for $\chi_1$ and $\chi_3$ class PCA spectra in the $2.5-25$~keV energy band 
with the normalization for TCAF frozen at the values as obtained from solely TCAF model fits over a smaller energy range. We have 
listed the variations of TCAF model parameters, viz. disk accretion rate ($\dot{m}_d$), halo rate ($\dot{m}_h$), shock location ($X_s$), 
the shock strength ($R$) and the additional cutoffpl model parameters with the chosen time segments along with the respective 
error bars. The reduced $\chi^2$ of the spectral fits are also listed in the last column.}
{\centerline{}}
\begin{center}
\begin{tabular}{c@{\hskip 0.1in} c@{\hskip 0.1in} c@{\hskip 0.1in} c@{\hskip 0.1in} c@{\hskip 0.1in} c@{\hskip 0.1in} c@{\hskip 0.1in} c@{\hskip 0.1in} c@{\hskip 0.1in} c@{\hskip 0.1in} c@{\hskip 0.1in} c}
%\begin{small}
\hline
\hline
\rm Variability & $ \rm \dot{m}_d $ & $ \dot{m}_h $ & $ X_s $ & $ R $ & 
 $\alpha$ & $\beta$ & cutoffpl & $\chi^2/dof$ \\
 class  & (in $\dot{M}_{Edd}$) & (in $\dot{M}_{Edd}$) &  (in $r_s$) & &  & (in keV) & Norm  & \\
 \hline
\vspace{0.2cm}
           
            $\chi_1$ &   $2.494_{-0.02}^{+0.04}$ & $0.367_{-0.002}^{+0.002}$ &  $32.434_{-0.10}^{+0.11}$ & $1.052_{-0.005}^{+0.006}$ & $1.871_{-0.04}^{+0.04}$ & $13.674_{-0.02}^{+0.02}$ & $6.323_{-0.04}^{+0.04}$ & $35.66/42$ \\            
\vspace{0.2cm}
            $\chi_3$ & $1.131_{-0.01}^{+0.02}$ & $0.172_{-0.002}^{+0.002}$  & $21.064_{-0.14}^{+0.14}$ & $1.373_{-0.004}^{+0.005}$ & $1.500_{-0.02}^{+0.02}$ & $15.064_{-0.08}^{+0.08}$ & $2.744_{-0.01}^{+0.01}$ & $40.48/41$\\
 \hline
 \vspace{0.2cm}
 \end{tabular}
\end{center}
\end{table}

\vskip 0.1cm
\noindent {\bf $\bullet$ Spectral fitting using \textit{TCAF+cutoffpl} model for $\chi_1$ class:} 
The continuous 3000 second observation was chosen for spectral fitting in the $2.5-25$~keV energy range. The disk rate and 
the TCAF normalization over the entire $2.5-25$~keV range were found to be around 1.75 $\dot{M}_{Edd}$ and 17.5 respectively. 
The spectral fitted parameters using TCAF+cutoffpl model in the 2.5-25.0 keV have been given in Table 2. The $\alpha$ and 
$\beta$ parameters in cutoffpl model have been found to be around 2.08 and 15.95 keV respectively. 
However, as stated earlier, because of the presence of outflow, the entire range could not be fitted with TCAF solely. 
In case of $\chi_1$ class, only in the range 2.5-16.5 keV, the spectral fitting could be accomplished using TCAF only. 
The normalization was found to be around $27.46$.

The TCAF+cutoffpl fitting was repeated by keeping TCAF normalization frozen at $N=27.46$ and the flow parameters were 
extracted. The parameters did not change significantly from those reported in Table 1. The fitted parameters have been 
listed in Table 3. In figure 5(b), the data to model ratio along with the individual components is given. 

\begin{figure}[h!]
\vskip 0.1cm
\begin{center}
\includegraphics[scale=0.4,angle=0,width=8.5truecm]{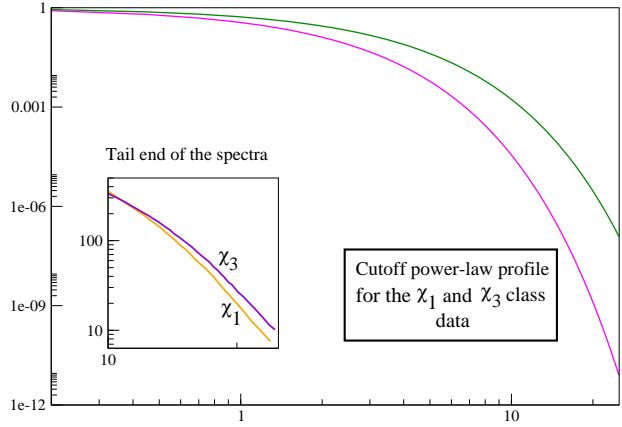}
\caption{Comparison of the cutoffpl profiles as obtained from the fitted parameters and tail end of the actual spectra. 
The relative hardness of $\chi_3$ class as found from the fitted parameters is in agreement with the actual observation.}
\end{center}
\label{fig4}
\end{figure}

\vskip 0.1cm
\noindent {\bf $\bullet$ Spectral fitting using \textit{TCAF+cutoffpl} model for $\chi_3$ class:} 
We wanted to study the relative behaviour of $\chi_1$ and $\chi_3$ class in terms of flow parameters. For that purpose, 
the same kind of spectral analysis as that of $\chi_1$ class have been accomplished in $\chi_3$ class as well. In this case, 
continuous 2000 second observation span had been chosen for spectral fitting using TCAF plus cutoff power-law (cutoffpl) model 
in the $2.5-25$~keV range. The disk rate and the TCAF normalization over the entire $2.5-25$~keV range had been obtained to be 
around $1.13$ $\dot{M}_{Edd}$ and $2.85$ respectively. These are relatively lower than that of $\chi_1$ class, which was in 
agreement with the higher low energy flux in case of $\chi_1$ class (Fig. 3). However, in case of $\chi_3$ class since the 
sub-Keplerian accretion rate reduces compared to $\chi_1$ class, the shock location moves inwards. Compared to $\chi_1$ class, 
the lower $\alpha$ and higher $\beta$ in case of $\chi_3$ class was in conjunction with the relative hardness of $\chi_3$ 
class data. The fitted parameters are listed in Table 2.

However, $\chi_3$ class was harder, the spectral fitting using only TCAF could be attained in the $2.5-14$~keV range. The disk rate 
obtained in this case was found to be lower as compared to $\chi_1$ class. The normalization was found to be around $5.55$. The 
normalization for TCAF was frozen at this value and spectral fitting was repeated over the entire 2.5-25.0 band. The rest of the 
parameters were not found to be changing significantly from those obtained from the parameters with normalization kept free. 
The cutoffpl parameters were still in conformity with the relative hardness of $\chi_3$ class. The lower disk rate for $\chi_3$ class 
accounting for smaller low energy flux was ensured in this case as well. The fitted parameters are listed in Table 3. In figure 6(b), 
we provide the 1,2 and 3$\sigma$ contours of the parameters $\dot{m}_d$ and $\dot{m}_h$.

\begin{figure}[htb]
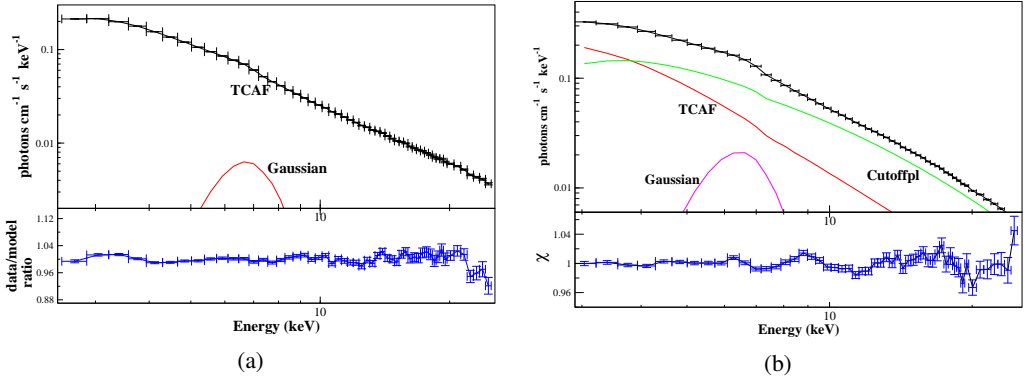

    \centering
\begin{subfigure}{0.45\textwidth}
  \includegraphics[width=\linewidth]{fig5a.eps}
  \caption{}%(a)}
  \label{fig:5a}
\end{subfigure}\hfil
\begin{subfigure}{0.45\textwidth}
  \includegraphics[width=\linewidth]{fig5b.eps}
  \caption{}%(c)}
  \label{fig:5b}
\end{subfigure}\hfil
\caption{(a) The data to model ratio for a fit in the $\chi_2$ class data using \textit{TCAF} model. A Gaussian is added to take 
into account the iron line emission $\sim$ 6.5 keV. (b) The data to model ratio for fitted spectrum of $\chi_1$ class data using 
\textit{TCAF + cutoffpl} model. The Gaussian profile for iron K$\alpha$ line emission is added.}
\label{fig5}
\end{figure}

\begin{figure}[htb]
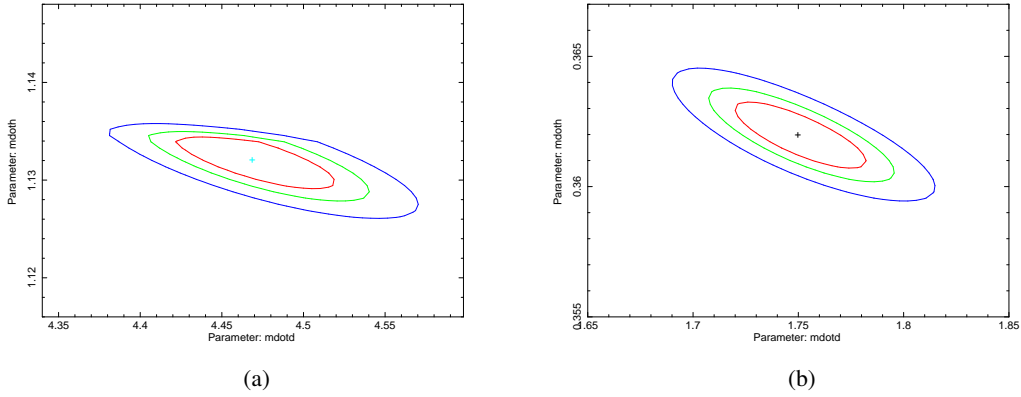

    \centering
\begin{subfigure}{0.50\textwidth}
  \includegraphics[angle=270,width=\linewidth]{fig6a.eps}
  \caption{}%(a)}
  \label{fig:6a}
\end{subfigure}\hfil
\begin{subfigure}{0.50\textwidth}
  \includegraphics[angle=270,width=\linewidth]{fig6b.eps}
  \caption{}%(c)}
  \label{fig:6b}
\end{subfigure}\hfil
\caption{The 66-90-99$\%$ confidence contours for the fitted parameters $\dot{m}_d$ and $\dot{m}_h$ in case 
of $\chi_4$ and $\chi_1$ classes respectively are shown in (a) and (b).}
  \label{fig6}
\end{figure}

\section{Summary and Concluding Remarks}
Depending upon the nature of the variation in intensities, hardness, and other distinct spectral and timing properties, 
light curves of GRS~1915+105 were classified into $14$ classes. IGR~J17091-3624 is another BH, which belongs to this special 
category of class objects and it also showed some of these class of light curves. The $\chi$ class, which has $4$ different ($\chi_{1-4}$) 
subclasses. In general, transient BHs show these class of light curves in hard and hard-intermediate spectral states. 
These classes generally show steady variation of intensities or fluxes. But we should mention one of distinct differences 
of these four $\chi$ classes, such as jet or outflow nature. $\chi_{2,4}$ classes are radio-quiet where as $\chi_{1,3}$ classes 
are radio-loud. 
In this paper, we concentrated on the $\chi$ class data and carried out spectral fits of all the four subclasses using 
the additive table model fits file, generated using Two Component Advective Flow solution of Chakrabarti \& Titarchuk (1995) 
after some necessary modifications (for more details, see Debnath et al. 2014, 2015a). We attempted to provide of a plausible 
picture of the accretion flow dynamics. Given that in $\chi$ class observations of GRS 1915+105, the X-ray flux showed 
negligible fluctuation and the spectra strikingly resemble the hard states of BHBs, and given also that TCAF solution 
has been successful in explaining the spectral behaviour of transient black holes in extracting physical flow parameters from fits, 
we were interested in looking at the spectral behaviour of $\chi$ class data of the variable source GRS 1915+105. This is the 
first time that such a variable source data is fitted with the TCAF model. The theoretical backdrop for the explanation of the 
time variability of this kind of variable sources under the purview of two component advective flow was envisaged by Chakrabarti 
and his collaborators (Fig. 1(a,b)). There it has been stated that the manifestation of all the different classes could be a 
consequence of the variation of the local accretion flow parameters, the emergence of outflow under suitable conditions, the 
collapse of outflow under sufficient cooling within the sonic radius ($\sim 2.5~X_s$) and local enhancement of accretion flow 
because of return of a portion of the outflow (Chakrabarti et al., 2000). Among them, those flow configurations in which the 
flow is steady and outflow is absent; or those which are endowed with a steady outflow would be associated with a steady flux. 
They are identified as the $\chi$ classes, which we considered for our analysis in this paper. 
In case of $\chi_{2,4}$ classes which are devoid of outflow, the spectral fitting are accomplished with TCAF+Gaussian model only. 
Here, additional Gaussian was used to take of the Iron emission line $\sim 6.5$~keV. Throughout the analysis, we have not kept the mass of the black hole as a free parameter. This is because of the fact that mass is an intrinsic property of the black hole and not a flow parameter. Since we are interested in the variation of flow parameters across different subclasses, we decided to peg the mass at a known value. The mass have been kept frozen at $14M_\odot$ as estimated in Greiner et al. (2001). This falls within the ballpark of more recent estimation of $12.4_{-1.8}^{+2.0}~M_\odot$ by Reid et al (2014). We observe that flow parameters do not significantly alter even if we change the mass $\sim 15\%$ within the observational ballpark. Even if we fix the mass $\sim 12.4M_\odot$, the accretion rates change only $\sim$ 0.5\% - 0.8\%. Therefore, it does not affect our conclusions.
\par
The low energy flux ($<$ 10 keV) was significantly 
greater in case of $\chi_4$ class. The disk rate of the $\chi_4$ class turned out to be more as compared to that of $\chi_2$ class 
since the disk photons primarily contribute to the lower energy end of the spectra. However, since the outflow is not present, 
the normalization between these classes are almost the same (see, Table 1).\par
If we compare the spectral nature of $\chi_2$ and $\chi_4$ classes, we observe a higher photon energy flux in both soft and hard domains in case of $\chi_4$ class (Figure 2). The ratio of energy flux in harder (10.0-25.0 keV) and softer (2.5-10.0 keV) domains turns out to be $\sim$  0.32 and 0.63 in case of $\chi_2$ and $\chi_4$ class respectively. Therefore, the flux of high energy photons appears to be more in case of $\chi_2$ class. However, the ratio of halo accretion rate and disk accretion rate in case of $\chi_2$ and $\chi_4$ turns out to be 0.20 and  0.25 respectively. This apparent paradox is resolved when one notes that the flux does not depend only on the accretion rates, but on all the flow parameters, such as the shock location and shock strength. For this reason, the ratio of fluxes between two classes can not be equated with just the accretion rate ratios. If we had a controlled system where one accretion rate can be changed keeping other parameters constant then such comparisons would have made sense. In our case, the shock location decreases from $\sim 47 ~ r_s$ to $\sim 38 ~ r_s$ as we move from $\chi_2$ to $\chi_4$ class. Shock strength also does not remain constant. Higher shock location and stronger shock in case of $\chi_2$ class naturally intercepts more number of soft photons and consequently produce more hard photons through inverse Compton scattering. All such factors contribute to the overall flux and hardness of the spectra. However, If we compute the contribution of TCAF in the overall energy flux in 2.5-25.0 keV, we observe that in case of $\chi_4$ class the flux is greater by a factor of $\sim 4.8$ compared to $\chi_2$ class. This result is in agreement with Figure 2.
\par 
The launching of outflow from within the CENBOL region would be sensitive to the flow parameters. In has been shown in C99 that outflows from CENBOL are possible in harder states. The process is not sensitive to the total accretion rate, but the location of the shock and the compression ratio $R$ at the CENBOL boundary. Often high accretion rates quench the jet base and reduces the thermal pressure required for the launching of jets. In case of $\chi_{1,3}$ class the outflow is collimated and steady. Consequently, the inverse Comptonization of the seed photons from the base of outflow is present and that is manifested as an additional power-law. We had to employ cutoffpl model in addition 
to TCAF for best spectral fitting. The two parameters $\alpha$ (the power-law photon index) and $\beta$ (the exponential roll-off factor) 
dictates the strength of the outflow. From the comparison of the spectral nature of the two classes, the relative dominance of low energy
flux for the $\chi_1$ class data and also its overall relative hardness are apparent. These two features must be reflected 
in the spectral fitted parameters. In the $\chi_{1,3}$ class data, since the cooling within the sonic sphere is not catastrophic and 
there is no fall back of the outflow on the initial accreting matter, the fitted parameters must not also change over the period of 
observation. These are the features that were actually obtained as has been reported earlier. The disk rate as obtained in case of 
$\chi_1$ class data was dominating over $\chi_3$ class. The lower $\alpha$ and higher $\beta$ of $\chi_3$ class data accounted 
for the relative hardness of that class. The work done by Rao et al. (2000) resolved the $\chi_3$ class data into multi-colour 
disk black-body, a Comptonized component and a power-law by fitting the spectra with diskbb+CompST+cutoffpl model. 
Since the disk and Comptonized component from the CENBOL have been already incorporated in the TCAF solution, therefore only 
the addition of cutoffpl model along with the TCAF solution should be sufficient to account for the contribution from 
the outflow. The result we have obtained thus vindicated our expectation. However, in order to segregate the contributions 
from the two components, we did the spectral fitting in two phases. Using TCAF solely, the spectra were fitted in a smaller 
energy range. Subsequently, the spectral analysis were repeated in the entire $2.5-25.0$ keV energy range with TCAF+cutoffpl model 
keeping the TCAF normalization frozen at the average value from the earlier fitting. The values of the fitted parameters did not change 
significantly, and the essential relative features between the spectra of $\chi_1$ and $\chi_3$ class data still remained. 
Thus the one-to-one correspondence of the $\chi_{1,3}$ class of the object with the accretion flow configuration as depicted 
in Figure 1(a,b) is justified, with the effect of Comptonization being absorbed in the cutoffpl model. The relative dominance 
of outflow in case of $\chi_{3}$ class data is well corroborated from the $\alpha$ and $\beta$ parameters.\par 

From Figure 5(b), we observe the cutoffpl contribution starts dominating over TCAF above $\sim$ 4 keV. However, this does not make TCAF redundant anyway. Cutoffpl is added to account for the power-law contributed by inverse Comptonized photons from outflow base. The outflow, on the other hand, is launched from within the CENBOL region, which, in turn, is produced because of the interplay of accretion rates. Therefore, the accretion flow and outflow together form a complete system. Therefore, the description of the system can be complete only if both of these segments are described by respective models. The cutoffpl parameters are indicative of the relative strength of the outflow in different subclasses, which is depicted in Figure 4.
\par 

Earlier all the classes of this interesting source GRS 1915+105 were characterized by the Comptonizing Efficiency (CE), which is nothing but the ratio of power-law and black body photons (Pal, Chakrabarti \& Nandi, 2001). This is different from the standard hardness ratio because the energy domain corresponding to soft and hard photons dynamically vary from one class to another.  This is a measure of geometry variation and how the CENBOL is evolving as the source transits from one class to another.  Therefore, the sequencing of CE in ascending order provides us a plausible sequence of transition of the source from one variability class to another. On the other hand, spectral fitting of the observed spectra using TCAF provides us a description of the accretion flow around the source in different classes in terms of basic flow parameters. In case of TCAF, the soft and hard fluxes are not calculated separately. Rather, all the flow parameters contribute to fit the observed spectral energy distribution. For this reason, the individual flow parameters or their ratios can not be connected to the CE of respective classes. Rather, all the flow parameters are important for a meaningful description of each individual subclasses.  Also, incorporation of associated phenomenological models like cutoff power-law can provide us information regarding strength of the outflow present in different subclasses. Therefore, spectral fit using physical models TCAF provides us a deeper understanding of the characteristics of different subclasses in terms of accretion flow dynamics. In Banerjee et al. (2020), TCAF has already been applied to study the flow parameters in case of $\theta$ class data of GRS 1915+105.

%#### 
The next logical step would be to embark upon the spectral analysis on intermediate classes that depict wide range of time variabilities. 
We believe that the root of such variabilities is the changing configurations of the accretion flow and during the transition from burst-off 
to burst-on state the accretion rates and the normalization would change accordingly to suit that feature. This process is underway and 
the results would be reported elsewhere.

\normalem
\begin{acknowledgements}
A.B. and A.B. acknowledge support fellowship of S. N. Bose National Centre for Basic Sciences, Kolkata, India. 
D.D. acknowledges partial support of the DST/GITA sponsored India-Taiwan collaborative project fund (GITA/DST/TWN/P-76/2017).
S.K.C. and D.D. acknowledge partial support from ISRO sponsored RESPOND project (ISRO/RES/2/418/17-18) fund. 
Research of D.D. and S.K.C. is supported in part by the Higher Education Dept. of the Govt. of West Bengal, India. We would also like to thank the anonymous referee, whose comments and suggestions have contributed to enhance the quality of this paper.
\end{acknowledgements}

%\section{REFERENCES}

\end{document}